\documentclass[aps,prd,twocolumn,preprintnumbers,superscriptaddress,dblfloatfix,nofootinbib]{revtex4-1}
\usepackage[utf8]{inputenc}
\usepackage{lmodern}
\usepackage{amsmath,amssymb,mhchem}
\usepackage{graphicx}
\usepackage{url}
\usepackage{color}
\usepackage{enumitem}
\usepackage{subfigure}
\usepackage[dvipsnames]{xcolor}
\usepackage[colorlinks=true,breaklinks=true]{hyperref}
\hypersetup{allcolors=[rgb]{0.0 0.0 0.6},linkcolor=[rgb]{0.75 0.05 0.05}}
\usepackage{bm}
\usepackage{epsfig}
\usepackage{amsmath}
\usepackage{amssymb}
\usepackage{slashed}
\usepackage{color}
\usepackage{accents}
\usepackage{url}
\usepackage[dvipsnames]{xcolor}
\usepackage{cancel}
\usepackage[normalem]{ulem} % to add \xout

\newcommand{\exclude}[1]{}
\newcommand{\Sec}[1]{Sec.~\ref{#1}}

\newcommand{\Fig}[1]{Fig.~\ref{#1}}
\newcommand{\Tab}[1]{Table~\ref{#1}}

\begin{document}

\preprint{IPMU21-0005}

\title{Exploring the Origin of Supermassive Black Holes with Coherent Neutrino Scattering}

\author{V\'ictor Mu\~noz} 
\email{victor.manuel.munoz@ific.uv.es}
\affiliation{Instituto de F\'isica Corpuscular (IFIC), CSIC-Universitat de Valencia, \\ Apartado de Correos 22085, E-46071, Spain}

\author{Volodymyr Takhistov} \email{volodymyr.takhistov@ipmu.jp}
\affiliation{Department of Physics and Astronomy, University of California, Los Angeles \\ Los Angeles, California, 90095-1547, USA}
\affiliation{Kavli Institute for the Physics and Mathematics of the Universe (WPI), UTIAS \\The University of Tokyo, Kashiwa, Chiba 277-8583, Japan}

\author{Samuel J. Witte} 
\email{s.j.witte@uva.nl}
\affiliation{Instituto de F´ısica Corpuscular (IFIC), CSIC-Universitat de Valencia, \\ Apartado de Correos 22085, E-46071, Spain}
\affiliation{Gravitation Astroparticle Physics Amsterdam (GRAPPA), Institute for Theoretical Physics Amsterdam and Delta Institute for Theoretical Physics, University of Amsterdam, Science Park 904, 1098 XH Amsterdam, The Netherlands}

\author{George M. Fuller}
\email{gfuller@ucsd.edu}
\affiliation{Department of Physics, University of California, San Diego, La Jolla, CA 92093-0319, USA
}

\date{\today}

%%%%%%%%%%%%%%%%%%%%%%%%%%%%%%%%%%%
\begin{abstract}
Collapsing supermassive stars ($M \gtrsim 3 \times 10^4 M_{\odot}$) at high redshifts can naturally provide seeds and explain the origin of the supermassive black holes observed in the centers of nearly all galaxies. During the collapse of supermassive stars, a burst of non-thermal neutrinos is generated with a luminosity that could greatly exceed that of a conventional core collapse supernova explosion. In this work, we investigate the extent to which the neutrinos produced in these explosions can be observed via coherent elastic neutrino-nucleus scattering (CE$\nu$NS). Large scale direct dark matter detection experiments provide particularly favorable targets. We find that upcoming $\mathcal{O}(100)$ tonne-scale experiments will be sensitive to the collapse of individual supermassive stars at distances as large as $\mathcal{O}(10)$ Mpc.
\end{abstract}
%%%%%%%%%%%%%%%%%%%%%%%%%%%%%%%%%%% 

\maketitle

\tableofcontents

%%%%%%%%%%%%%%%%%%%%%%%%%%%%%%%%%%%

\section{Introduction}

In this paper we explore the utility of using coherent elastic neutrino-nucleus scattering in dark matter experiments for detecting the neutrinos produced in the collapse of supermassive stars to black holes. Two issues are at the heart of why detecting these neutrinos is problematic: (1) Unlike in conventional massive star core collapse, the neutrinos generated in the collapse of a supermassive star are relatively lower energy, reflecting their thermal origin, and making them hard to detect; and (2) As yet there is no direct observational evidence for the existence of supermassive stars. However, new exploration of this subject is called for, first because the mystery of the existence of supermassive black holes at high redshift continues to deepen, and second because experimental techniques have dramatically improved, leading to the first detection of coherent elastic neutrino-nucleus scattering in 2017~\cite{Akimov:2017ade}.

Supermassive black holes (SMBHs) with masses $ \sim10^6 - 10^9 M_{\odot}$ are thought to be ubiquitous in the centers of galaxies~\cite{vanderMarel:1997hr}, and serve as the central engines powering quasars and Active Galactic Nuclei (AGN)~\cite{Rees:1984}. The existence of $10^9 M_\odot$ SMBHs at redshifts as high as $z \sim 7$ presents an intriguing astrophysical problem, as both Eddington-limited accretion and successive mergers are challenged in growing $\mathcal{O}(M_\odot)$ black holes to these masses on the relevant timescales~\cite{Mortlock:2011,Venemans:2013npa,Wu:2015,Banados:2017unc}.

Many proposals have been put forth to explain the origin and formation mechanism of SMBHs~(e.g.~\cite{Kroupa:2020}, see Ref.~\cite{Volonteri:2010} for review). 
Of the standard astrophysical pathways to SMBH formation~\cite{Begelman:1978}, 
several go through an intermediate supermassive star (SMS) with mass $M \gtrsim 3 \times 10^4 M_{\odot}$. Large black holes would be the likely result of the collapse of such SMSs. In turn, these black holes would act as seeds~\cite{Woods:2018lty}. Through accretion or mergers, these could grow into SMBHs. There is no compelling argument for the existence of such SMSs, and no direct observation of them. However, they could plausibly arise either from a primordial gas cloud or as a consequence of the evolution of a dense star cluster. All we can say for certain is that such a configuration, should it arise, will collapse via the Feynman-Chandrasekhar general relativistic instability once it is primarily supported against gravitation by components moving at or near the speed of light, photons in the case of SMSs, and stars in the dense star cluster case. SMS collapse to a black hole will be accompanied by a prodigious neutrino burst, with luminosities capable of exceeding conventional core-collapse supernova by several orders of magnitude~\cite{Shi:1998nd}. 

The physics accompanying the collapse of SMSs has been extensively studied in a variety of environmental conditions, including accretion and rotation~(e.g.~\cite{Fuller:1986,Baumgarte:1999nh,Saijo:2002qt,Umeda:2016smj,Haemmerl:2017,Haemmerl:2018,Nagele:2020xqq}).~These events are expected to produce an array of experimental signatures (e.g.~\cite{Shapiro:1979,Fuller:1986,Sun:2017}), including the generation of gravitational waves~\cite{Shibata:2016vzw,Uchida:2017qwn,Li:2018}, gamma-ray bursts~\cite{Fuller:1997em,Sun:2017}, and neutrinos~\cite{Shi:1998nd,Shi:1998jx,Linke:2001mq}.

Detection of neutrinos from SMS explosions would provide invaluable information regarding SMBH seed formation. In contrast to standard core-collapse supernovae, SMS neutrinos would be produced with an energy spectrum generated by the annihilation of thermal $e^\pm$ pairs, and that is similar among the various emitted neutrino species. Note, however, that the $\nu_e$ and $\bar\nu_e$ fluxes will be larger than those of the $\mu$ and $\tau$ flavor species~\cite{Shi:1998nd} because of the charged current annihilation channel available for production of electron flavor neutrinos. The neutrinos produced via thermal $e^\pm$-pair annihilation could be detected either directly from the collapse of individual relatively nearby objects, or via the diffuse background produced from the cumulative history of SMS collapses (see e.g.~\cite{Shi:1998jx,Shi:1998nd}). In the latter case, the spectrum may suffer significant redshift, causing the entirety of the spectrum to become buried under the large neutrino fluxes generated by the Sun, reactors, nuclear processes in the Earth, etc. The possibility of detecting neutrinos from SMS explosions through inverse beta decay (IBD) $\overline{\nu}_e + p \rightarrow n + e^+$ has been previously considered~\cite{Shi:1998nd,Shi:1998jx}, both with conventional neutrino telescopes, such as Cerenkov-based Super-Kamiokande~\cite{FUKUDA:2003,Abe:2013gga}, and with IceCube~\cite{Ahrens:2002dv}. 

Coherent elastic neutrino-nucleus scattering (CE$\nu$NS) could provide a new way to search for the low energy neutrinos of a SMS collapse-generated neutrino burst. In contrast to IBD, CE$\nu$NS is unconstrained by the IBD kinematic threshold on neutrino energy. Moreover, CE$\nu$NS will have sensitivity to all six $(\nu_{e}, \overline{\nu}_e, \nu_{\mu}, \overline{\nu}_{\mu}, \nu_{\tau}, \overline{\nu}_{\tau})$ neutrino flavors. CE$\nu$NS has been recently directly observed~\cite{Akimov:2017ade}, and has been considered in a range of studies related to neutrino physics, including
sterile neutrinos (e.g.~\cite{Pospelov:2011ha,Billard:2014yka}),
non-standard neutrino interactions (e.g.~\cite{Harnik:2012ni,Dutta:2017nht,Flores:2020lji}),
solar neutrinos (e.g.~\cite{Billard:2014yka,Schumann:2015cpa,Franco:2015pha,Gelmini:2018ogy}),  
geoneutrinos~\cite{Gelmini:2018gqa}, 
neutrinos from dark matter (DM) annihilation and decays~\cite{Cherry:2015oca,Cui:2017ytb,McKeen:2018pbb}, as well as supernova~\cite{Chakraborty:2013zua,XMASS:2016cmy,Lang:2016zhv,Kozynets:2018dfo,Khaitan:2018wnf} and pre-supernova neutrinos~\cite{Raj:2019wpy}.

Large scale direct detection experiments, whose primary target is dark matter (DM) observation, are themselves effective neutrino telescopes and can explore complementary parameter space compared to that of conventional neutrino experiments. In particular, such experiments have very low keV-scale thresholds, potentially providing sensitivity to a complementary part of the neutrino spectrum. Furthermore, with heavy nuclei as detector targets, these experiments are particularly well suited for signal detection via CE$\nu$NS, whose cross-section scales approximately as neutron number squared.

In this study we explore the detection capabilities of large scale direct DM detection experiments via CE$\nu$NS of neutrinos produced from SMS collapse. We examine both the signal arising from the collapse of individual objects, as well as the diffuse signal generated by the cumulative collapse rate throughout their history. The former of these could be detectable from collapsing stars in nearby galaxies.

This work is organized as follows. In \Sec{sec:sms} we describe neutrino production from the collapse of supermassive stars. \Sec{sec:dd} presents an overview of large direct detection experiments and their sensitivity to coherent neutrino scattering. The sensitivity of these experiments to an individual explosion of a supermassive star and the diffuse background is presented in \Sec{sec:signal}. In \Sec{sec:darkmatter} we elaborate on the extent to which supermassive star collapses may contribute to the background of dark matter searches. We conclude in \Sec{sec:con}.

\section{Neutrinos from Supermassive Star Collapse}\label{sec:sms}

\subsection{Neutrino burst}\label{sec:sms}

Supermassive stars with masses $M \gtrsim 3 \times 10^4 \, M_\odot$ are expected to directly collapse into a black hole as a result of the Feynman-Chandrasekhar instability, unless centrifugal forces from rapid rotation or magnetic fields are sufficiently strong~\cite{Fuller:1986}. During the collapse, only a fraction of the initial star, the homologous core (HC) comprising $M_5^{\rm HC}/M_5^{\rm init} \simeq 0.1$ of the initial mass, with $M_5$ being stellar mass in units of $10^5 M_{\odot}$, plunges through the event horizon, resulting in prompt black hole formation. Most of the HC binding energy will be trapped within the BH, but a small fraction could be emitted in the form of neutrinos and (an even smaller fraction) in gravitational waves. 

Neutrino emission from SMS collapse has been analyzed in Ref.~\cite{Shi:1998jx,Shi:1998nd}, whose discussion we follow. The entropy-per-baryon in SMS is large, corresponding to low density and modestly high temperature, with electromagnetic equilibrium consequently implying a large electron-positron ($e^\pm$) pair density. Neutrino pairs are produced by $e^+e^-$ annihilation in the in-falling HC, with most of the neutrino luminosity being generated as the radius of the star nears trapped surface formation, its Schwarzschild radius. Unlike core collapse supernovae, the in-falling material is transparent to emitted neutrinos. Consequently, the luminosity, spectrum, and time profile are well-defined quantities\footnote{Numerical hydrodynamic simulations of Ref.~\cite{Linke:2001mq} show emission suppressed by up to two orders compared to the analytic results of Ref.~\cite{Shi:1998nd}. These differences stem from differing treatments of the in-fall and collapse timescales, pressure,
and the adiabat of collapse, and are exacerbated by the $T^9$ dependence of the neutrino emissivity. Significant uncertainties remain. We employ the results of Ref.~\cite{Shi:1998nd} throughout this study as an example (for comparison of models see Fig. 3 of Ref.~\cite{Li:2018})}. In particular, the total neutrino luminosity is expected to be a sizable fraction of the HC gravitational binding energy $E_s \simeq 10^{59} M_5^{\rm HC}$~erg. Reference \cite{Li:2018} showed that the SMS HC mass range that gives an optimal fraction of the rest mass radiated as neutrinos is $5\times{10}^4\,{M_\odot} < M^{\rm HC} < 5\times{10}^5\,{M_\odot}$.  Other factors that determine the ultimate neutrino fluence from collapse of these objects include the time profile of the collapse, dictated by a number of features\footnote{The in-fall time can increase in the presence of rotation or strong magnetic fields.} of the HC. Roughly, this collapse time scale will be $t_s \simeq M_5^{\rm HC}$~s. The neutrino energy spectra and fluxes are determined mostly by the evolution of the density and temperature distributions near the Schwarzschild radius (i.e. a rapid rise as the mass in-falls, followed by a rapid fall as material is absorbed by the black hole).  

\begin{figure}[t]
\begin{center}
\includegraphics[width=1\columnwidth]{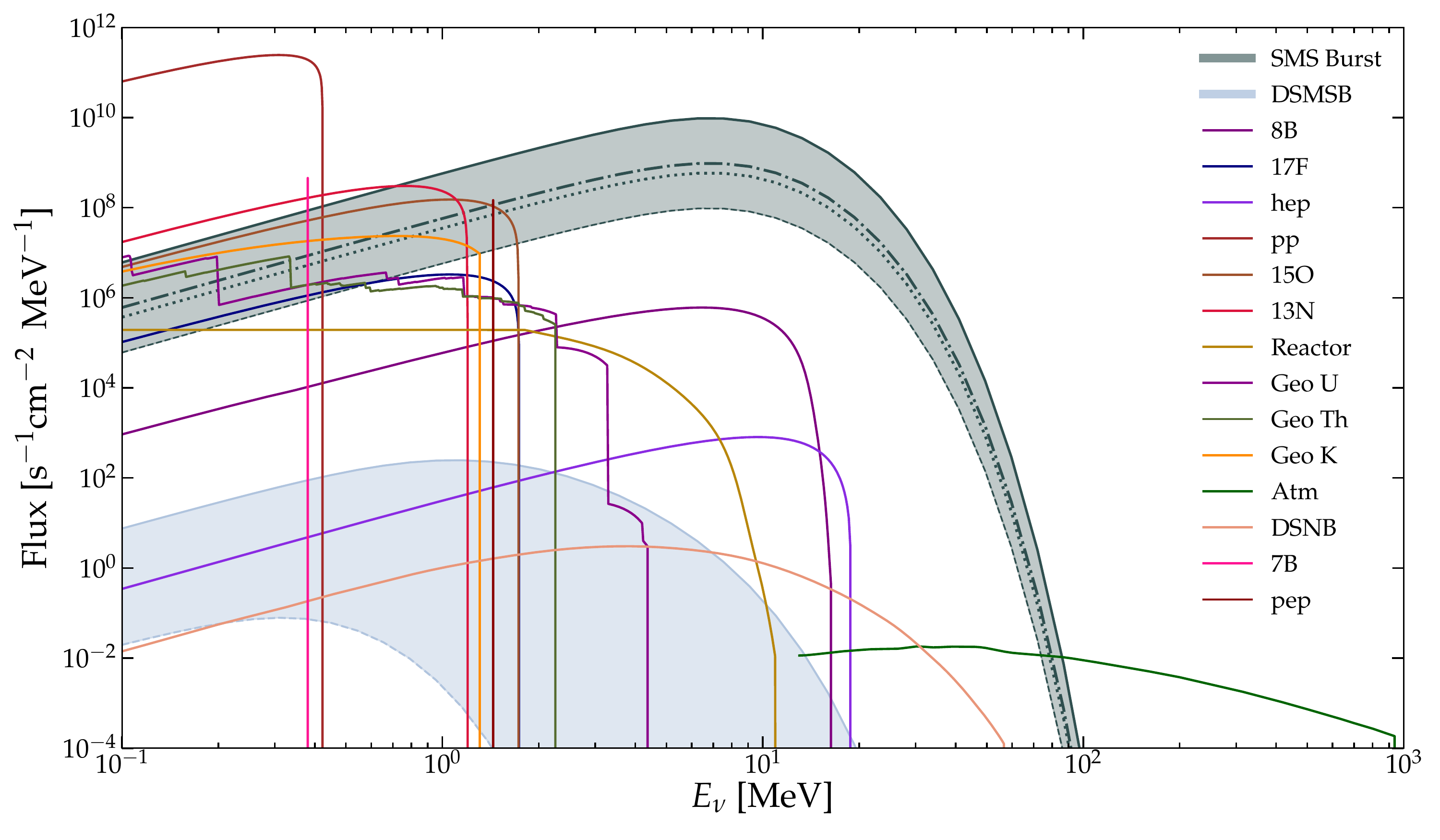} 
\caption{Neutrino flux from a single SMS explosion (black) and the diffuse background generated from the cosmological evolution of SMS collapse (DSMSB, blue). The  band  detailing  the burst signal assumes a SMS with mass $10^{5}M_{\odot}$ collapsing at 100 kpc (solid), at 1 Mpc (dashed), at 1 Mpc with enhancement due to non-negligible rotation or magnetic field (dash-dotted) and a SMS with mass $3\times10^{4} \,M_{\odot}$ exploding at a distance of 1 Mpc (dotted). The  band  detailing  the diffuse background signal corresponds to the $R^{\rm flat}$ (upper) and $R^{\rm Pop3}$ (lower) models. Shown for comparison are the solar ($^8$B,$^{17}$F,$^{15}$O,$^{13}$N,$^7$Be,hep,pp,pep), geo ($^{238}$U,$^{232}$Th,$^{40}$K), reactor, atmospheric neutrino and diffuse supernovae neutrino background (DSNB) fluxes -- reactor and geoneutrino fluxes have been computed assuming the experiment is located at SNOLAB~\cite{Gelmini:2018ogy}. }
\label{fig:Nu_flux}
\end{center}
\end{figure}

Considering the peak emission occurring near the Schwarzschild radius, the resulting neutrino luminosity from pair-production during SMS star collapse can be estimated as~\cite{Shi:1998nd}
\begin{equation}
     L_\nu \simeq 3 \times 10^{57} \, (M_5^{\rm HC})^{-1.5} \, {\rm erg \, cm^{-3} \, s^{-1}}~.
\end{equation}
The associated neutrino spectrum can be well-fit by
\begin{equation}
    f_\nu(E) \simeq \left(\frac{1.2 \, {\rm MeV}}{ \sqrt{M_5^{\rm HC}}}\right)^{-3} \, \frac{1}{F_2(2)} \frac{E^2}{e^{(E \sqrt{M_5^{\rm HC}} / 1.2 \, {\rm MeV}) - 2}+1}~,
\end{equation}
where 
\begin{equation}
    F_k(\eta_\nu) = \int_0^\infty \, \frac{x^k \, dx}{e^{x-\eta_\nu} +1} \, 
 \end{equation}
 and the average neutrino energy is $\left<E_\nu \right> \simeq 4 \, (M_5^{\rm HC})^{-1/2}$ MeV.
 
The presence of a strong magnetic field or rapid rotation will delay the SMS collapse. Neutrinos produced near the Schwarzschild radius will then have a higher chance of escaping before the core moves through the event horizon. This results in the possibility of an increase in the emitted neutrino fluence by up to an order of magnitude, and an increase in neutrino energies by a factor of two compared to the case of a non-rotating and non-magnetized collapse scenario~\cite{Shi:1998nd}. The partition of energy among the neutrino species remains the same, however.

Despite the enormous neutrino luminosities from SMS collapse, the detection of this signal on large cosmological scales is unlikely~\cite{Shi:1998nd}. However, detection prospects are favorable if the redshift of SMS collapse event is $z \lesssim 0.2$ (i.e. $\sim 1$~Gpc distance). Since there exist many quasars and AGN at these redshifts, the collapse rate could be sufficiently high so as to be within reach of detection.

 \subsection{Diffuse neutrino background}\label{sec:sms}
 
 An isotropic background of redshifted neutrinos will be generated by the cosmological history of SMS collapse. Given the complete ignorance of the formation and collapse rate of such large stars, we adopt a phenomenological perspective in which we motivate a variety of different redshift-dependent collapse rates, and investigate the detection prospects for each. 

 The flux of diffuse neutrinos from SMSs can be computed from the neutrino emission spectrum and the collapse rate $R_{SMS}(z)$ via
 \begin{equation}\label{eq:DNB}
     \frac{d\phi}{ dE_{\nu}}(E_{\nu}) = \int \, dz \, \frac{R_{SMS}(z)}{H(z)} \, f_{\nu}( E_{\nu} (1+z))  \, ,
 \end{equation}
where we adopt cosmological parameters consistent with the latest \textit{Planck-2018} measurements~\cite{Akrami:2018vks}. 

\begin{figure*}[tb]
\centering
\begin{subfigure} 
 \centering
 \includegraphics[width=0.45\linewidth,trim={0 0 0  0},clip]{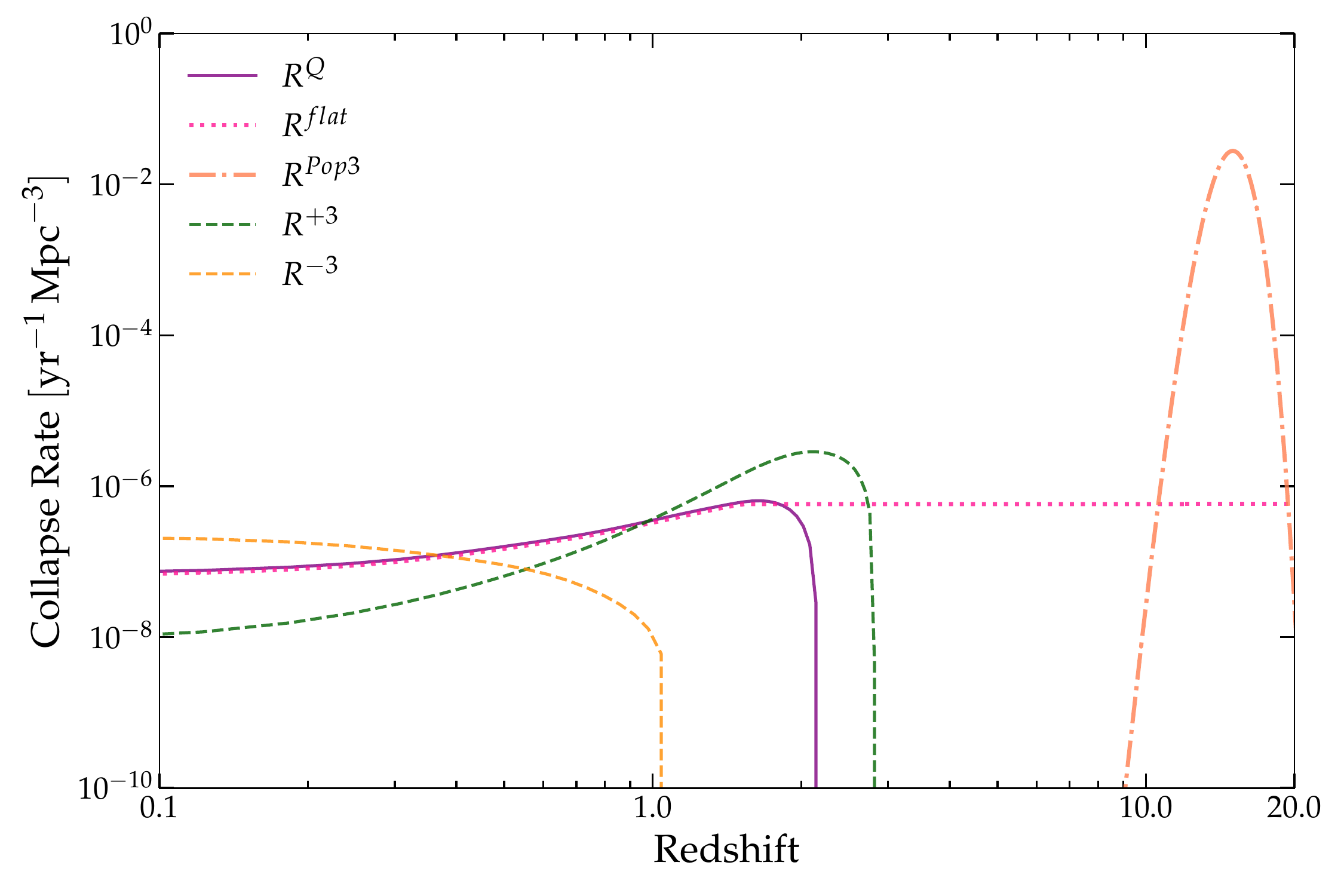}
\end{subfigure}
\begin{subfigure} 
 \centering
 \includegraphics[width=0.45\linewidth,trim={0 0 0 0},clip]{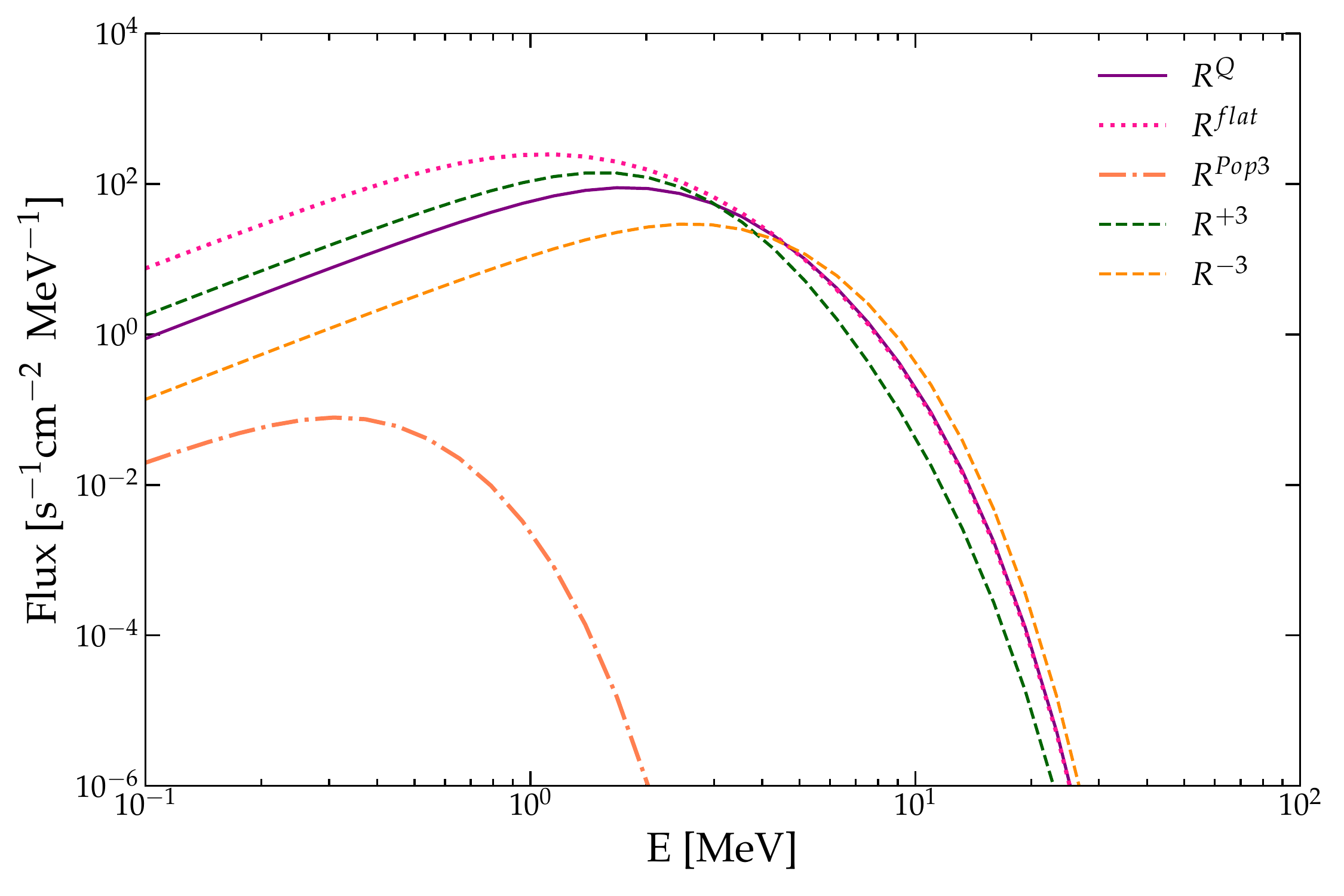} 
\end{subfigure}
\caption{Historical SMS collapse rate [Left] and the resulting diffuse neutrino flux at Earth [Right] for each of the models considered in \Sec{sec:sms}. }
\label{fig:collapse}
\end{figure*}

In what follows, we adopt five different parameterizations of the collapse rate in order to obtain a rough estimation of uncertainty in the flux and spectral shape. The models assume
\begin{enumerate}
\item The collapse rate of SMSs traces the quasar formation rate. If we assume the typical quasar lifetime (which is much shorter than the Hubble time) is redshift-independent, then we can assume that the formation rate directly follows the quasar number density. We take this rate to be consistent with the results of~Ref.~\cite{shaver1996decrease,veron1998eso}, and call this model $R^Q$.
\item In order to asses the impact of additional redshift-dependent factors not directly included in the quasar formation rate, we consider two models in which $R^Q$ is re-scaled by a factor of $(1+z)^\alpha$. In order to understand extreme variations in this factor, we adopt $\alpha = \pm 3$, and denote each model by $R^{\pm 3}$. 
\item As will be shown, $R^Q$ decreases dramatically at redshifts $z \sim 2$. Should SMSs be the origin of SMBHs, the collapse rate must extend to much larger redshifts. To account for this, we adopt a model which is consistent with the quasar formation rate at $z \leq 1.5$, and is flat at $1.5 \leq z \lesssim 20$, the upper cut-off taken to be roughly consistent with the onset of star formation. We call this model $R^{\rm flat}$.
\item Finally, we adopt a model in which SMSs are assumed to form predominately in metal-free environments at high redshifts. It has been suggested that low-metallicity environments could allow for the rapid cooling and formation of such objects, implying a preferential formation rate peaking near $z \sim 15$. We model this using a Gaussian distribution centered at $z = 15$ with a width $\Delta z = 1$. We call this model $R^{Pop3}$, as it would suggest these stars are among those first produced in the Universe (i.e. Pop-III stars, or perhaps stars produced by tidal disruption in dense star clusters). 
\end{enumerate}
The aforementioned models are assumed to characterize only the redshift dependence of the collapse rate. In order to determine normalization of the rate, we assume that less than $10\%$ of the baryons have resided in SMSs. That is, we define the baryon density in SMSs to be 
\begin{equation}
\rho_{SMS}= \int dt\, M \, \frac{R_{SMS}(z)}{(1+z)^{3}} \, .
\end{equation}
If these black holes do indeed serve as the seeds for supermassive black holes at the center of galaxies, it would be reasonable to estimate that approximately one SMS exists per galaxy, or equivalently $\rho_{SMS} \sim (M / 10^{10} M_\odot) \rho_b$. We use this to normalize the SMS collapse rate, and show the resultant histories, and the subsequent neutrino fluxes as observed here at Earth, in \Fig{fig:collapse}.

\section{Large Direct Detection Experiments} \label{sec:dd}

\subsection{Experimental configurations}

\begin{table}[t]
\centering
\begin{tabular}{c|c|c|l} 
 \hline
  \hline
 Target  &  Mass  &  Threshold  &  Reference \\ 
     &  (tons) &   (keV)&   \\ 
 \hline
 Ar  & 300 & 0.6 & ARGO~\cite{zuzel2017darkside,Aalseth:2017fik} \\ 
 Xe & 50 & 0.7 & DARWIN~\cite{Aalbers:2016jon,Aprile:2016wwo} \\
 Pb & 2.4 & 1.0 & RES-NOVA~\cite{Pattavina:2020cqc} \\ 
 \hline
  \hline
\end{tabular}
\caption{Considered detector configurations.}
\label{tab:exp_conf}
\end{table}

In this study, we consider detector configurations consistent with the proposed specifications of the upcoming direct DM detection experiments DARWIN~\cite{Aalbers:2016jon,Aprile:2016wwo}, using xenon (Xe) as a target material, and ARGO~\cite{zuzel2017darkside,Aalseth:2017fik} using argon (Ar) as a target material. These experiments are able to achieve considerable fiducial volume while also taking advantage of a keV-level energy threshold. In addition, we also consider\footnote{We note that low-background xenon and argon detectors have been in development for many years and the scalability of these setups has been established. The feasibility of Pb-based detector on a competitive scale is still to be demonstrated.} a configuration based on lead (Pb), following the recently proposed RES-NOVA~\cite{Pattavina:2020cqc} experiment for detection of core-collapse SN neutrinos via CE$\nu$NS. An overview of these configurations is listed in \Tab{tab:exp_conf}.
 
We assume the experiments are located at SNOLAB
(Sudbury, Canada), which is likely to host a number of next-generation direct detection experiments.
We stress, however, that this assumption does not strongly affect our conclusions.
The depth of this lab (6010 m.w.e.) ensures that backgrounds due to cosmogenic muons are highly suppressed.
 
Throughout this work, we will optimistically consider that experiments have perfect detection efficiency and energy resolution, and we adopt detection thresholds consistent with the targeted low-energy searches of each experiment. Furthermore, when considering neutrino coherent interactions with the nuclei, the expected background is assumed to arise exclusively from other neutrino sources\footnote{This assumption is in principle not fully realistic as, e.g., the ionization signal ``S2-only'' analyses of argon and xenon~\cite{Aprile:2019xxb} have unavoidable electronic backgrounds (but allow for lower signal thresholds). However, since the SMS burst signal occurs over a period of $\sim \mathcal{O}(1 {\rm s})$, time correlations should easily allow one to differentiate this signal from background.}. This assumption allows us to treat all analyses on an equal footing, and provide general results independent of specific configurations that could change in the future. 

\subsection{Scattering rates}

Given a neutrino flux $\phi_{\nu} (E_{\nu})$ the resulting differential event rate per unit time and detector mass as a function of the recoil energy $E_R$, per unit time and mass $m_I$ of a target nuclide $I$ in a detector is given by
\begin{equation} \label{eq:nu_diff_rate}
\dfrac{d R_{\nu}^I}{d E_R} = \dfrac{C_I}{m_I} \int_{E_{\nu}^{\rm min}} \phi_{\nu} (E_{\nu}) \dfrac{d \sigma^I (E_{\nu}, E_R)}{d E_R} d E_{\nu}~,
\end{equation}
where $d \sigma^I (E_{\nu}, E_R) / d E_R$ is the coherent neutrino-nucleus scattering differential cross-section and $C_I$ is the fraction of nuclide $I$ in the material. In case several nuclides are present, individual contributions are summed.

For a target mass $m_I$ at rest, the minimum neutrino energy required to produce a recoil of energy $E_R$ is
\begin{equation}
E_{\nu}^{\rm min} = \sqrt{\dfrac{m E_{R}}{2}}~.
\end{equation}
The maximum recoil energy due to a collision with a neutrino of energy $E_{\nu}$ is
\begin{equation}
E_R^{\rm max} = \dfrac{2 E_{\nu}^2}{m + 2 E_{\nu}}~.
\end{equation}

\begin{figure*}[tb]
\centering
\begin{subfigure} 
 \centering
 \includegraphics[width=0.45\linewidth,trim={0 0 0  0},clip]{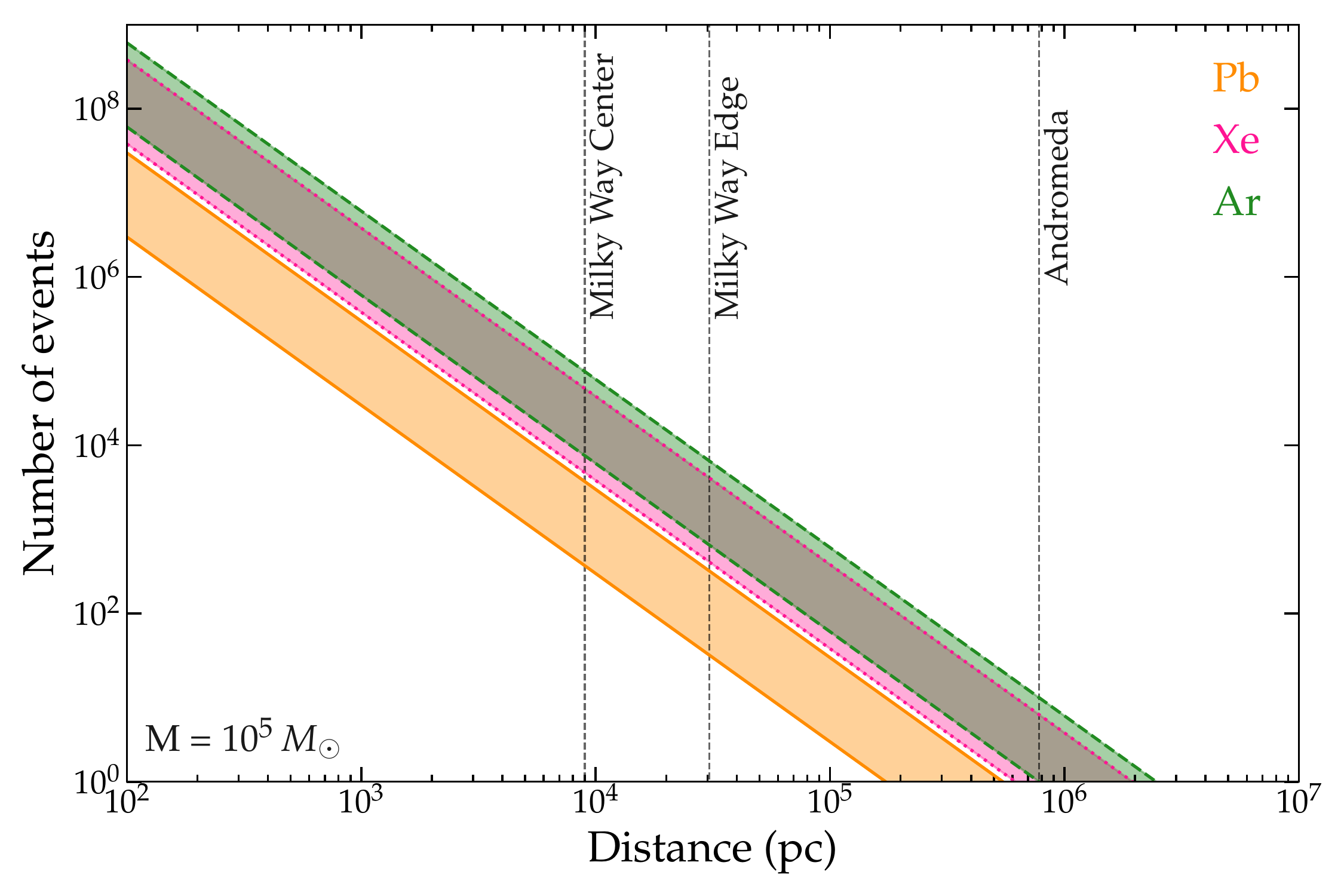}
\end{subfigure}
\begin{subfigure} 
 \centering
 \includegraphics[width=0.45\linewidth,trim={0 0 0 0},clip]{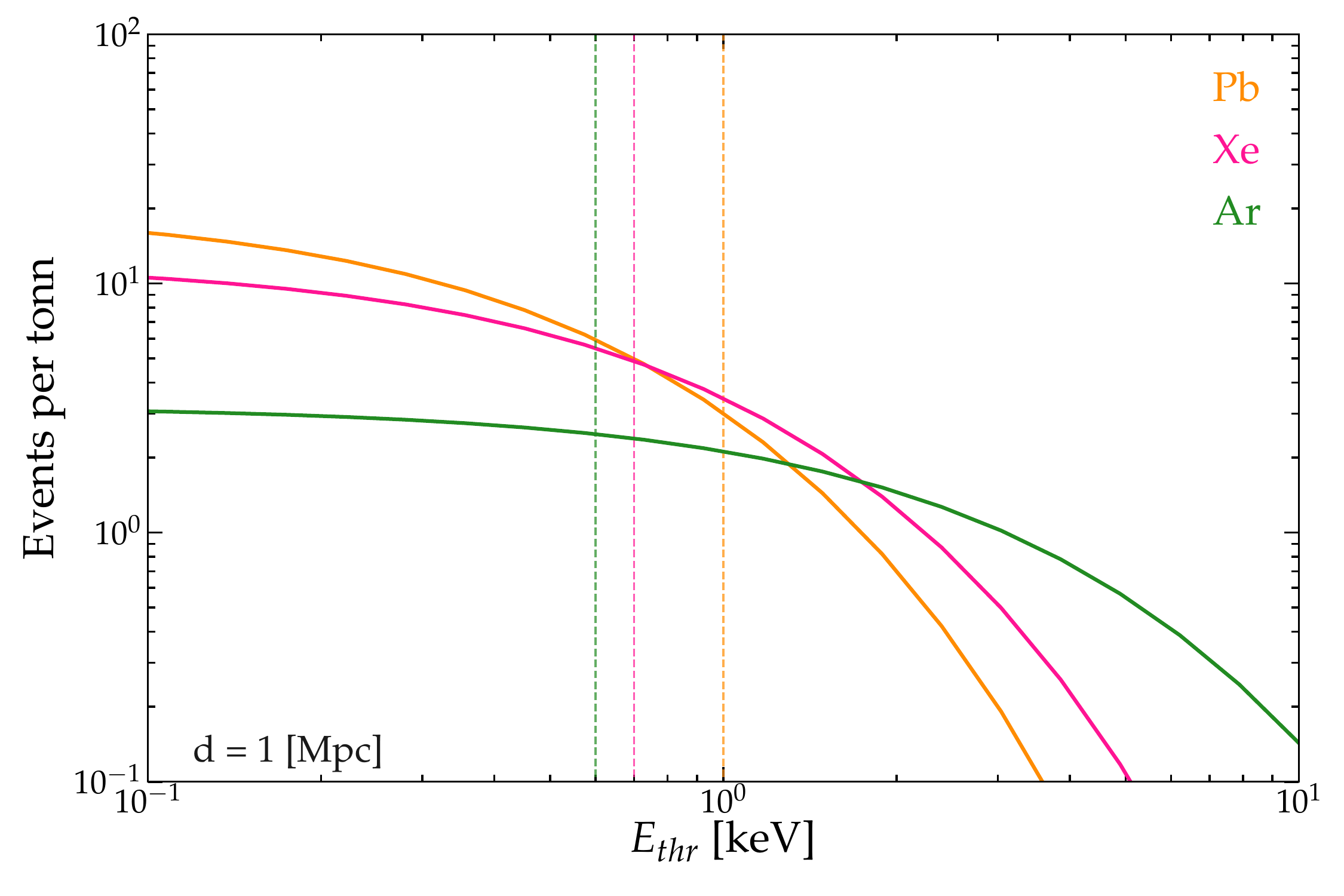} 
\end{subfigure}
\caption{[Left] Number of events detected from the burst of a SMS with mass $M = 10^{5} M_{\odot}$ for each experimental configuration as a function of distance. The shaded regions indicate the potential enhancement in the signal that may arise if the SMS has a non-negligible rotation or magnetic field. Shown for comparison are distance markers denoting the location of the galactic center, the edge of the Milky Way, and Andromeda. [Right] Number of events per tonne of detector mass for each target element as a function of threshold energy $E_{th}$, computed assuming $M = 10^5 \, M_\odot$ and $d = 1 $ [Mpc]. The adopted experimental thresholds are shown with vertical lines.}
\label{fig:burstevents}
\end{figure*}

\subsection{CE$\nu$NS}

The Standard Model coherent-scattering neutrino-nucleus cross-section is given by~\cite{Freedman:1977xn} 
\begin{equation}\label{eq:nucxsec}
\dfrac{d \sigma^I (E_{\nu}, E_R)}{d E_R} = \dfrac{G_f^2 m_I}{4 \pi} Q_w^2 \left(1 - \dfrac{m_I E_R}{2 E_{\nu}^2}\right) F_{I}^2 (E_R)~,
\end{equation}
where $m_I$ is the target nuclide mass, $G_f$ is Fermi coupling constant, $F_{I}(E_R)$ is the form factor, which we take to be the Helm form factor \cite{Helm:1956zz}, $Q_w = [(1 - 4 \sin^2 \theta_{\rm W}) Z_I-N_I]$ is the weak nuclear charge, $N_I$ is the number of neutrons, $Z_I$ is the number of protons, and $\theta_{\rm W}$ is the Weinberg angle. Since $\sin^2 \theta_{\rm W} = 0.223$ \cite{Patrignani:2016xqp}, the coherent neutrino-nucleus scattering cross-section follows an approximate $N_I^2$ scaling.

\section{Supermassive Star Neutrino Signal Detection} \label{sec:signal}

In Fig.~\ref{fig:Nu_flux} we depict the expected neutrino flux for SMS collapse at a distance of 0.1-1 Mpc and with varying HC mass and inclusion of rotation/magnetic fields. In Fig.~\ref{fig:burstevents} we illustrate the expected number of events from the collapse of a SMS as a function of explosion distance. In this case, we illustrate the enhancement effect (shaded band) that may arise should the SMS star rotate or have strong magnetic fields.  Fig.~\ref{fig:burstevents} shows the event rate normalized by the fiducial volume as a function of the detection threshold, highlighting that lead and xenon will benefit particularly from lowering the detection threshold. Note that the adopted thresholds are shown with the colored vertical lines.

\begin{figure}[tb]
\centering
 \includegraphics[width=1\columnwidth]{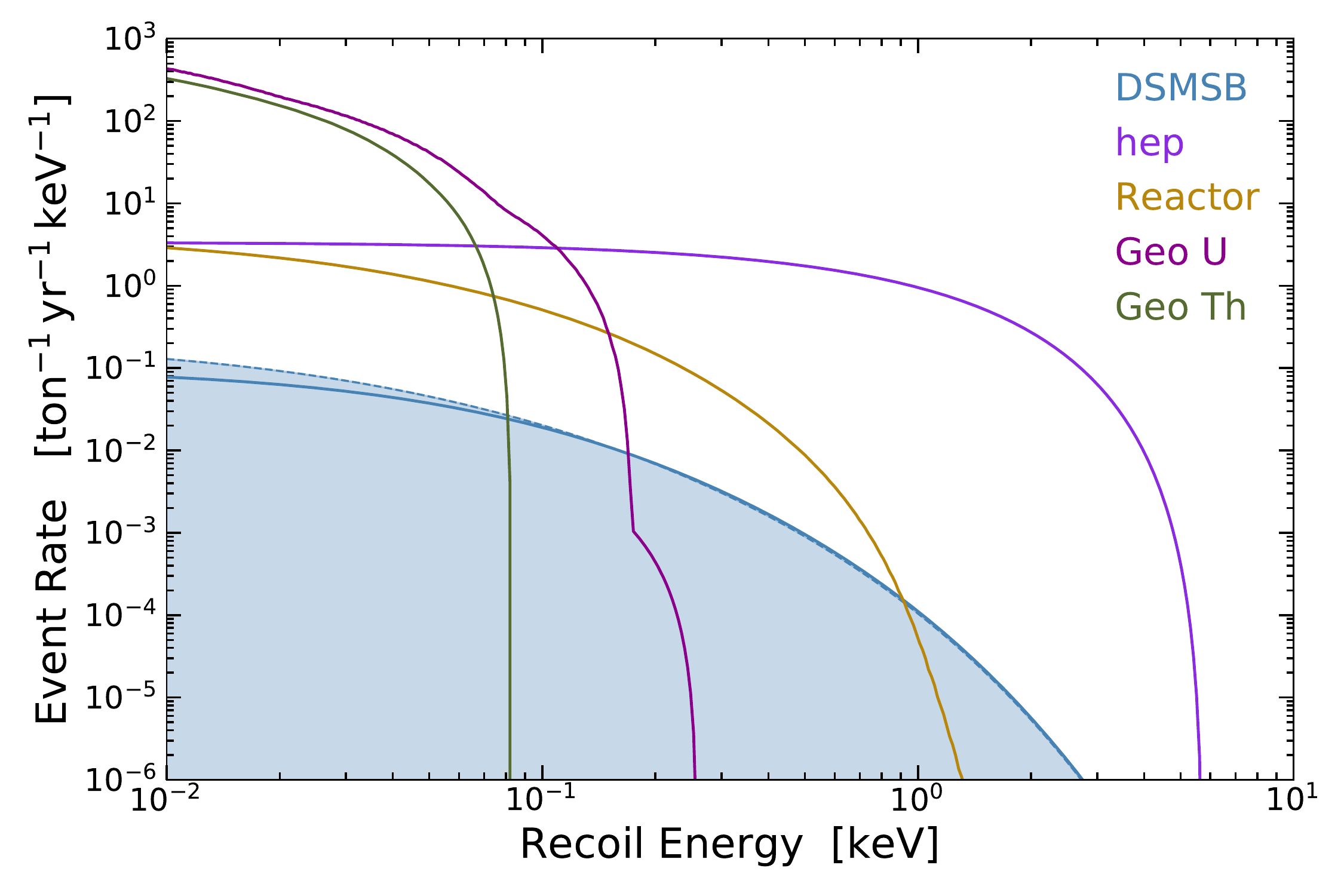} 
\caption{Detection of diffuse supermassive  star  neutrino  background (DSMSB), assuming a xenon target and experiment located at SNOLAB. DSMSB signal is shown for $R^Q$ model (solid) and the $R^{\rm flat}$ model (dashed), with shaded region extending down to the signal detection obtained for the  $R^{\rm Pop3}$ model. Contributions for solar (hep), geo ($^{238}$U,$^{232}$Th) and reactor neutrinos are displayed (see Ref. \cite{Gelmini:2018ogy} for details). }
\label{fig:diffuse}
\end{figure}

In Fig.~\ref{fig:diffuse} we illustrate the event rate produced in a xenon-based experiment by the diffuse SMS neutrino background. Various background neutrino sources are shown for comparison. We expect no more than one event will be detected using the experimental configurations listed in \Tab{tab:exp_conf}, implying that it will be a difficult task to disentangle the diffuse background from the other neutrino sources. Nevertheless, DSMSB will contribute to the irreducible background in the searches for dark matter. For neutrino sources with a well-defined spectrum and flux, this irreducible background may be partially circumvented via background subtraction techniques; this is not the case, however, for the diffuse neutrino flux from SMSs. We now turn our attention toward addressing the potential difficultly that could arise from such a background in the search for dark matter.
  
\section{Dark Matter and the Diffuse Neutrino Background} \label{sec:darkmatter}

Individual collapses of super-massive stars are unlikely to obstruct the search for dark matter, as they will typically generate multiple nuclear recoils within a time window of $t \lesssim \mathcal{O}({\rm few})$ seconds. The diffuse background on the other hand has no strong time correlation, and in analogy to the effect of the diffuse supernova background (DSNB), this will necessarily contribute to the irreducible background in the direct detection searches for DM. We stress, however, that this will likely be a sub-dominant effect to other backgrounds.

The extent to which CE$\nu$NS inhibits DM searches has been discussed extensively in the literature within the context of an irreducible neutrino background constituting a ``neutrino floor'' (e.g.~\cite{Billard:2013qya,Monroe:2007xp,Gelmini:2018ogy,OHare:2020lva}). This question is often posed in the following manner: What exposure is required in order for an experiment to identify a particular DM candidate (with a well-defined mass and scattering cross section) at the statistical confidence level of $X \sigma$ (where $X$ is often taken to be $3$)? For a particular model of DM, and for a fixed experimental exposure, this defines a ``discovery floor''. The extent to which this discovery floor scales with exposure is critically dependent upon the level of degeneracy between the recoil spectrum of DM interactions and neutrinos. 

The limitations on the DM discovery potential could prove rather difficult to quantify, because the associated SMS diffuse neutrino energy spectrum is determined by the assumed SMS collapse rate. In turn, this is a completely unknown function of redshift. As discussed previously, it is reasonable to conjecture that the SMS redshift-dependent collapse rate could be strongly related to the quasar and AGN formation rate. However, this need not be the case and possible deviations from such scaling can lead to significant differences in the shape of the resulting scattering rate within the experiments. Furthermore, the normalization of the SMS collapse rate contains only an upper limit, which we can estimate by ensuring no more than $\sim 10\%$ of the baryons have resided in SMS. 

\begin{figure}
\begin{center}
\includegraphics[width=1\columnwidth]{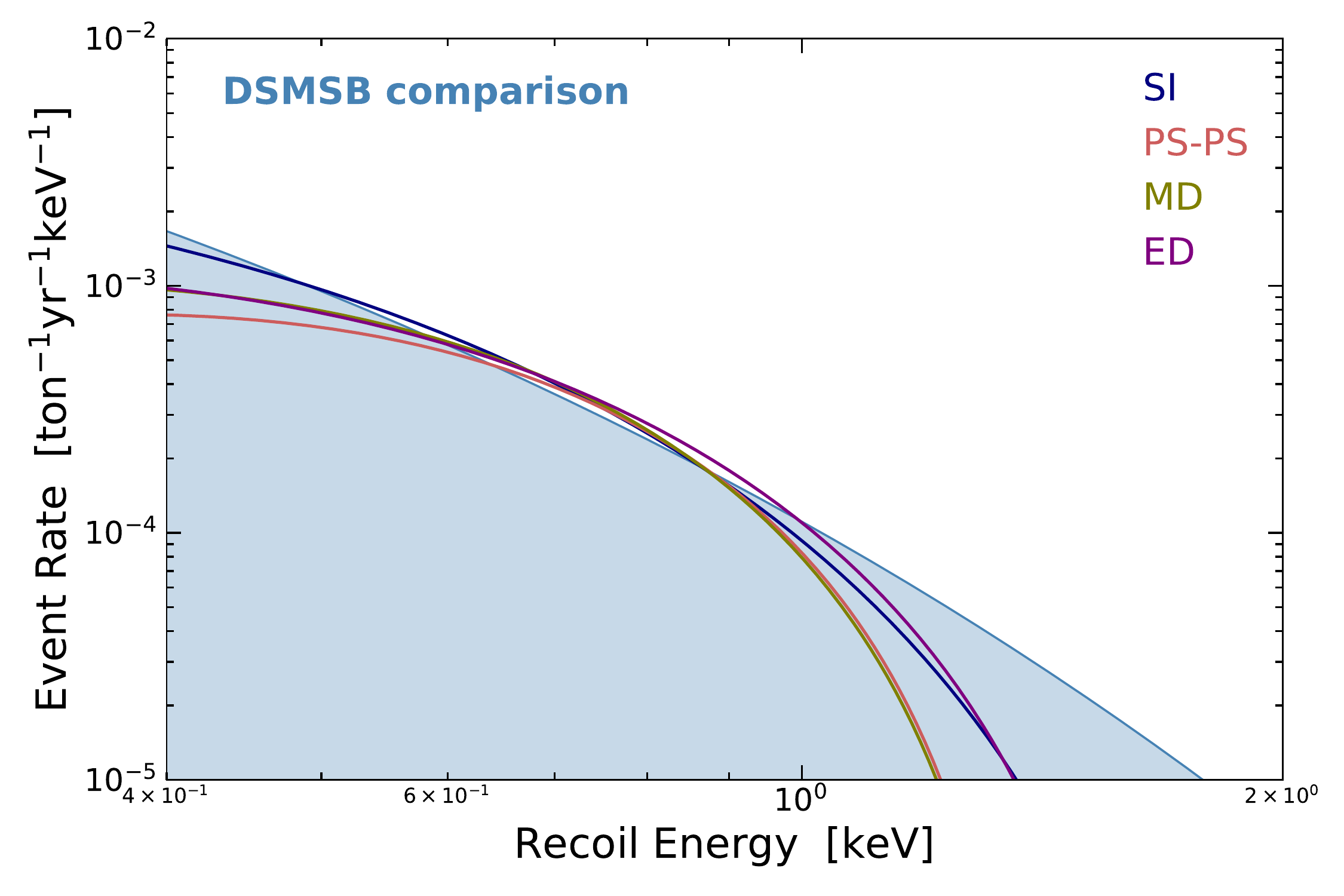} 
\caption{ Comparison of event rates for recoil spectra induced in a Xenon target from a DSMSB with the recoil spectra arising in various DM interaction models (see text). The best fit values for the DM parameters are given in table ~\ref{Tab:bf}.  }
\label{fig:bestfit}
\end{center}
\end{figure}

In order to demonstrate the impact of the diffuse SMS background, we plot in \Fig{fig:diffuse} a comparison of the nuclear recoil event rate produced by the diffuse SMS background and that from solar, geo, and reactor neutrinos. We show both the $R^Q$ model (blue, solid) and the $R^{flat}$ model (blue, dashed), and we shade down to the event rate produced by the $R^{Pop3}$ model (not shown). While the rate never exceeds those coming from known neutrino sources, it does become sizeable at low energies. In \Fig{fig:bestfit}, assuming for illustration the idealized scenario that other backgrounds can be subtracted or suppressed and SMS collapses pose the dominant background, we fit the event rate arising from the $R^Q$ model assuming dark matter interacts with nuclei through a spin-independent contact interaction (SI), a electric dipole (ED), a magnetic dipole (MD), or a pseudo-scalar contact interaction (PS) (see Ref.~\cite{Gelmini:2018gqa} for the details of each interaction). In~\Tab{Tab:bf} we display the approximate DM masses and cross-sections that would be recovered if the diffuse SMS background was mistakenly interpreted in the context of DM. We observe that the diffuse SMS background could further hamper the search for DM candidates with masses $m \lesssim 5$ GeV. Both \Fig{fig:bestfit} and the fit performed in \Tab{Tab:bf} assume that the other low energy neutrino backgrounds (e.g.~solar, reactor, geoneutrinos) can be effectively suppressed, which might be a challenging experimental task. Given both experimental and theoretical uncertainties (related \emph{e.g.} to the experimental response or the difficulties in computing the spectrum of other neutrino sources), the diffuse SMS background is unlikely to be the dominant inhibitor to the DM searches in the low energy regime. 

If SMS neutrino indeed eventually becomes a sizable effect in the context of DM searches, there are a number of possible ways in which this could be circumvented. First, experiments that run for multiple years could search for the annual modulation of the scattering rate, induced by the time-variation between the motion of Earth and DM rest frame. Uncertainties in the DM-nucleon interaction~\cite{DelNobile:2015tza,DelNobile:2015rmp} and the astrophysical distribution of DM~\cite{DelNobile:2015nua}, however, can significantly complicate the amplitude and phase of the annual modulation. In the absence of backgrounds, using the annual modulation to differentiate between DM models using only one detector typically requires $N_{\rm evts} \gg \mathcal{O}(10^3)$ events (neglecting astrophysical uncertainties) ~\cite{Witte:2016ydc}. With backgrounds, this number is likely orders of magnitude higher. Consequently, such a technique will not prove easy. A better understanding of halo uncertainties~\cite{OHare:2018trr,Buch:2019aiw,OHare:2019qxc} or the use of novel analysis methods~\cite{Gondolo:2017jro,Gelmini:2017aqe} may improve the situation. Alternatively, directional detection could allow one to efficiently remove isotropic backgrounds, leaving only the dark matter scattering rate~\cite{Mayet:2016zxu,OHare:2017rag,Vahsen:2020pzb}. 

\begin{table}[t]
\centering
\begin{tabular}{c|c|c} 
 \hline
  \hline
 Model   &  Mass (GeV) &  $\sigma$  (cm$^{2}$)   \\ 
     
 \hline
 SI  & 4.34 & 1.02 $\times$ $10^{-50}$  \\ 
 ED & 4.15 & 4.41 $\times 10^{-48} $ \\
 MD & 3.79 & 2.18 $\times 10^{-42} $ \\
 PS & 3.79 & 1.42 $\times 10^{-41} $ \\
 \hline
  \hline
\end{tabular}
\caption{ Best-fit mass and cross section for dark matter scattering with nuclei via a spin-independent (SI), electric dipole (ED), magnetic dipole (MD), or pseudoscalar (PS) interaction. Fits are to the $R^Q$ model of SMS collapse rate.}
\label{Tab:bf}
\end{table}

\section{Conclusions} \label{sec:con}  

Supermassive stars with mass $M \gtrsim 3\times 10^4 M_\odot$ are expected to directly collapse to black holes via the Feynman-Chandrasekhar instability. While no such stars have yet been directly observed, supermassive black holes at redshifts as high as $z \sim 7$ suggest at the least that initial seed black holes with large masses might be required. This also serves as a rationale for exploring the consequences of the existence of progenitor stars in this mass range at redshifts as early as $z \sim 15$. Should these objects exist, their collapse can yield a broad array of observable signatures, including gamma-rays, gravitational waves, and neutrinos. In this paper we have analyzed the extent to which neutrinos emitted from the collapse of such objects could be detected via coherent neutrino scattering, focusing on massive direct dark matter experiments. 
 
We have demonstrated that large scale underground experiments built for the purpose of detecting dark matter might be capable of identifying the collapse of individual supermassive stars in nearby galaxies, such as in Andromeda. A diffuse and isotropic neutrino background will also be produced from the cumulative historical collapse of such objects. We have analyzed a variety of potential redshift-dependent collapse rates that may arise, e.g., if the SMS collapse rate follows the AGN formation rate, or if SMSs are preferentially formed in metal-free environments, as would occur at higher redshifts (e.g. near $z \sim 15$). While we have focused on comparison of signal with other neutrino flux sources, future work on non-neutrino background suppression is essential for signal discrimination.
 
While the existence of SMSs has not been definitively established, such objects provide a simple and plausible explanation of the origin of the supermassive black holes observed to reside at the centers of galaxies, or a least the seed black holes needed to build them up by redshift $z\sim 7$. The only way to truly reveal the existence of these objects is to observe them. The neutrino flux produced from the collapse of SMSs offers a particularly intriguing channel in which to test their existence, as the neutrino energy spectra are non-thermal and easily distinguishable from other sources. Current direct dark matter experiments are already designed in a manner that is ideal for the search of such neutrino flux, with near-future experiments capable of probing the collapse of such objects on extra-galactic scales.
 
\acknowledgments
The work of S.J.W. was supported by a Juan de la Cierva Formacion fellowship, and is part of a project that has received funding from the European Research Council (ERC) under the European Union’s Horizon 2020 research and innovation programme (Grant agreement No. 864035 – UnDark). The work of V.M. was supported by CONICYT PFCHA/DOCTORADO BECAS CHILE/2018 - 72180000. The work of V.T. was supported by the U.S. Department of Energy (DOE) Grant No. DE-SC0009937. V.T. was also supported by the World Premier International Research Center Initiative (WPI), MEXT, Japan. G.M.F. and V.T. would like to thank Kavli IPMU, U. of Tokyo for hospitality where this work was initiated. G.M.F. acknowledges NSF Grant No. PHY-1914242 at UCSD and the NSF N3AS Physics Frontier Center, NSF Grant No. PHY-2020275, and the Heising-Simons Foundation (2017-228).

\bibliography{bibliography}
 
\end{document}